# Tidal Accelerometry: Exploring the Cosmos Via Gravitational Correlations

By

Timir Datta[1], Ming Yin[2], Mike Wescott[1], Yeuncheol Jeong [1], Pawel Morawiec [1],
James Gambrell [1], Dan Overcash [1], Huaizhou Zhang[1] and George Voulgaris[3]
1: Department of Physics & Astronomy
University of South Carolina, Columbia, SC 29208, USA
2: Physics and engineering Physics, Benedict College, Columbia, SC 29204, USA
3: Department of Earth and Ocean Sciences
University of South Carolina, Columbia, SC 29208, USA

## Overview

Gravitation – with its infinite range and the attractive only character – is not subject to shielding and all pervasive. Consequently the earth is immersed in a gravitational field of cosmic origin. In the Newtonian approach this field is described by a vector of the gravitational acceleration, **g**. It is a superposition of the fields due to all (N) sources at play but the direction of **g** need not lie along the line of sight to any source. At any location, the inclusion of additional sources can be enhance or reduce the net field. However, **g** being the physically relevant quantity neither non-linear ($N^2$) enhancement of classical interference nor HBT type, temporal anti-correlation associated with quantum interference ensues from such superposition.

The extra terrestrial field determines earth's own geodesic. But, on terrestrial objects its correlation in space is manifested as semi-diurnal and diurnal tides. Tides or geodesic deviation signals, $A(t)$, are a quasi-static, low frequency signal which arises from the relative changes in positions of the detector and source due to the rotation of the earth about its axis and is not part of the electromagnetic spectrum. In the Newtonian limit, the amplitude of $A$ is given by the product $G \cdot M \cdot \zeta \cdot R^3$, where $R$ is distance to mass $M$, $\zeta$ is baseline separation and G is Newton's constant and  can be directly calculated as a vector difference. These signals can be observed in a wide variety of natural and laboratory situations. Because $A(t)$ contains quantitative information we reason that it can be utilized to obtain quantitative astrophysical information. By a case study of a published report we show, how the systematic of the tidal signals can be used for calibration and de-trending which can significantly increase the confidence level of high precision observations.

Gravitational correlations can also be used to determine the distribution of celestial masses independently of the "1-2-3" law. Our estimates suggest that with a deflection sensitivity of $\delta R \sim 10^{-18}$ (m/s$^2$) it will be possible to resolve a diurnal signal of the purported 1 AU size, super-massive blackhole at the center of our galaxy, or a ten solar mass object at about a distance of 10pc. is

**1-Introduction**

Recently the world celebrated the 400[th] anniversary of Galileo and the optical telescope. Remarkably he was neither the inventor of telescopes nor the first person to use it in astronomy [1-4]. However, there can be little doubt about his genius in drawing the public's attention to the telescope which helped transform the 16[th] century curio into the versatile instrument it has become[5]. Also, Galileo's publication of "Sidereus Nuncius" in March, 1610 lead to the scientific revolution of the seventeenth century.

Even today, four centuries after Galileo and the invention of space crafts most of our understanding of the universe and outer space is based on information garnered by telescopes. For this purpose, a huge number of ground-based and space borne instruments are deployed round the clock. The radiation range that these modern detectors operate at has expanded immensely since Galileo and covers both waves and particles. Searches for the elusive gravitational waves are also in active progress [6-10]. However the basic "objective" has remained unchanged, the capturing of radiation that arrives to us from distant sources. Unfortunately, this modus operandi fails for objects that do not send out radiation, most well known example being dark matter objects.

The earth is also immersed in a two very well-known quasi-static fields, the first is geo-magnetic, fields that arise from sources that are terrestrial as well as those in outer space [11]. The scientific importance of magnetic observations was recognition in the early 19[th] century by Alexander von Humbolt the famous explorer and the great mathematician, Carl Friedrich Gauss. Currently a global network of magnetic observatories [12] provides geomagnetic field data, some of which extend back well into the 19th century, others give more recent observations at high acquisition rates, typically at a one-second average cadence. This resource is leading to a deeper understanding of the earth, those in the ionosphere, magnetosphere that extends hundreds of earth radii out into space, as well as to the conditions in the sun.

The other non-radiative field is gravitational and the subject of this article. This field determines the geodesic trajectory causing a test particle to undergo the familiar free fall acceleration of **g** towards the earth. However astronomical objects are also sources of gravity. A feature that makes gravitation extremely pervasive is unlike other interactions and radiations it cannot be shielded and hence even extra terrestrial gravitation penetrates equally into all media in every part of the globe [13].

Gravitational influences of celestial bodies produce perceptible effects on terrestrial systems. The most obvious example being sea or ocean tides. Water tides are so strong that they dominate human behavior over much of the coastal regions of the earth. Ocean tides which are principally of lunar and solar origins are extremely important in maritime activities, geology, marine, atmospheric and environmental sciences. Reproductive and migratory behaviors of many animals are synchronized with the tides. Etymology of the condition "lunacy" is another empirical evidence of celestial effects on human psychology. Furthermore, for a long time there has been at least anecdotal evidence, which is beginning to be scientifically substantiated about a credible connection between significant geophysical events such as earthquakes and tides [14].

For gravitational acceleration to be perceptible at a minimum correlations between two locations are required, this gives rise to a difference or tidal acceleration. This is due to the weak equivalence principle [15-18] in the Mach – Einstein sense that is an observer free falling inside a free falling elevator will have no sensation of being subjected to a force or accelerated. Or two apples following two nearby geodesics will remain parallel in empty flat space (fig. 1). But the apples will experience a tidal force and their paths tend to converge in a curved space-time, when approaching a massive object. So the curving of space indicates the presence of the mass.

At any point in space the static resultant Newtonian field, **g** is the linear vector superposition of the gravitational fields due to all the N individual sources and **g** is physically measurable quantity. Tides represent the vectorial difference between **g** at two locations separated by a baseline. It is not interference; in the sense, there is no non-linear ($N^2$) enhancement in the amplitude. Also there is no HBT [19] type quantum effect, i.e., increased tide at one detector does not correlate with lowered signal at another. Additionally its direction need not be along the baseline, or the line of sight to any of the sources. Reports of tidal influence in the laboratory relate to experiments that are designed to detect these effects as well as those where the signals show up incidentally in the data as modulations that correlate with astronomical predictions.

The pervasive nature of extraterrestrial gravitational influences in the laboratory is illustrated by the examples shown in figures 2- 4. The changes in the weight of an object with time are shown in figure 2; the data were obtained in our laboratory [20] with only minor software modifications on a commercial electronic balance. The influence of the astronomical forces on the test mass weight is indicated by the correlation between balance output in the laboratory in Columbia and the tidal water height at the port of Charleston, SC. The green rectangle highlights a period of the rapid changes in balance reading with time.

Next in figure 3 the tidal influences on the vertical free fall acceleration of a macroscopic test mass (bottom- shaded region) of an absolute gravimeter (solid blue curve) [21] and that of Cesium atoms in an atomic fountain [22] (top- solid red curve) are shown. Both of these papers [21,22] were designed to measure gravitational effects but tidal forces can also affect experiments that are not intended to involve "gravity"; a case in point is the changes in the lepton beam energy in the Large Electron-Positron Collider, LEP [23,24]. Figure 4 shows the beam energy dependence on the tidal variations the plotted data (solid red and blue dots) are of graphical accuracy as they were obtained from two separate publications; an estimated ~ 1ppm ($\Delta E/E$) decrease in beam energy per µ-gal increase in g is shown by the diagonal straight line (green).

Not surprisingly, the importance of this tidal signal as a source of astrometric information did not escape the notice of Isaac Newton [25-28]. He is the progenitor of the "1-2-3" law [29] as a method for determining the masses (M) of celestial objects from "satellite" time period and applied it to obtain accurate mass values of many objects in the solar system. Remarkably, it remains the method of choice for mass astrometry even to date [30]. However, this law is not applicable to objects without natural satellites, such as our Moon. Computation of the mass of the moon demands an alternative route to its gravitational field (acceleration). Coincidentally this impediment has been abated only recently by the advent of spacecrafts and artificial satellites [31-40]. In book 3, "System of the Worlds", of the Principia Sir Isaac [25-28] argued that of all the extant methods in the seventeenth century, only tidal changes were sensitive enough for the direct terrestrial determination of astronomical gravitational forces.

Although tidal changes in the sea level are associated with astrophysical forces but the forces are not faithfully measured by the actual water response. Because water height is profoundly more sensitive to the depth, size and shape of the sea basin than the vector force in question. Thus Newton's ($M_m/M_e$) ratio came out more than double the best current estimate (BCE) [41,42] of $1,230.00371\times10^{-2}$. This error placed the Earth-Moon barycenter outside the surface of the earth and was a concern amongst astronomers and experts [43-49] also impacted Newton's presentation of the later versions of Principia Mathematica and drew criticism. A massive moon would produce additional repercussions [50-53]. Here we reason that water tide clearly is not the way to go, nevertheless tidal or geodesic deviation signals can be an accurate source of astronomical quantities that may provide a double check on standard methods. More importantly gravitational tide can reveal information that cannot be accessed via conventional observational techniques.

In the following we will outline the vector computation of the relative gravitational acceleration or tidal signal. We show that with high resolution experimental data these results may be inverted to compute the masses and positions of gravitational sources in the sky.

## 2-Basics

The tidal vector $\Delta \boldsymbol{a}_{\text{cel}}$ is defined as the relative gravitational acceleration due to a celestial object, on a test object located at position "A" compared with that at "B" separated by a vector ζ. Hence, $\Delta \boldsymbol{a}_{\text{cel}}$ represents the relative correlation at two field points. It is a simultaneous measure of the gravitational field at the two locations A & B When the baseline ζ is small compared to other lengths (distances) in the problem then $\Delta \boldsymbol{a}_{\text{cel}}$ is the differential acceleration vector between two free-falling test particles following adjoining geodesics (fig.1).

In general relativity the relative acceleration of the separation between nearby geodesics relates to the Riemannian curvature, through the geodesic deviation equation as shown below [29]

$$\frac{d^2 \xi}{ds^2} = -R \xi$$

(2.1)

In the non-relativistic Newtonian limit geodesic deviation reduces to the difference in the gravitational fields at the two locations or the relative temporal acceleration between test particles. That is,

$$\Delta a_{cel} = a(R_A)_{cel} - a(R_B)_{cel}$$

(2.2)

Where, $\boldsymbol{a}(R_C)$ represents the gravitational acceleration produced by the celestial object at location C. The Newtonian gravitational field of the celestial body at a distant point $\boldsymbol{R}(t)$, or the acceleration of a small test mass $\boldsymbol{a}(\boldsymbol{R})_{\text{cel}}$ at time t, and is given by the familiar inverse square law formulae,

$$\vec{a}(\vec{R})_{cel} = GM_{cel}[\frac{\vec{R}(t)}{\lfloor R(t) \rfloor^3} \qquad (2.3)$$

Hence for a baseline vector $\zeta$, the $\Delta a_{cel}$ is given by,

$$\Delta \vec{a}(t)_{cel} = \vec{a}(\vec{R},t) - \vec{a}(\vec{R} - \vec{\zeta},t) \qquad (2.4)$$

For multiple objects, the total or observed signal $\Delta a$ is the vector sum of the $\Delta a_{cel}$ due to all the celestial mass in question. That is,

$$\Delta \vec{a}(t) = \sum \Delta \vec{a}(t)_{cel} \qquad (2.5)$$

### 3- Transformation of the acceleration vector between two inertial frames

Often it is necessary to consider the acceleration vector in two different frames of reference; in the following we will consider the relevant Euler transformations [54-56]. We consider the acceleration vector (**a**) of a free object in the laboratory (positioned in coordinate system **X'**, with origin at point marked *P*) relative to a geocentric inertial frame (**X**). At any given instant the inertial frame is co-moving with the earth and is in uniform motion with respect to a Newtonian frame (**X₀**) at rest with the distant stars. The system **X** is oriented such that the x-y lie on the equatorial plane of the globe; the x-axis points radially outward and passes the terrestrial equator at Greenwich meridian. The x'-y' plane of the surface centric system is the local horizontal plane but the axes are not oriented with the geographic cardinal directions (figure 5). The transformation from **X** to **X'** comprise of a rotation by the three Euler angles ($\phi, \theta$ & $\psi$) and a translation $r_e$ along z' to the point *P* on the surface as follows:

$$X = R(\phi,\theta,\psi)X + r_e \qquad (3.1)$$

In equation 3.1, $r_e$ is the radial distance between the laboratory and the center of the earth. If the coordinates of *P* are given by longitude ($\Lambda$), latitude ($\lambda$) and axis x' is at $\Psi$ (figure 2) degrees with east then the rotation matrix $R(\phi,\theta,\psi)$ is given by the following expression :

$$R(\phi,\theta,\psi) = \begin{bmatrix} R_{11} & R_{12} & R_{13} \\ R_{21} & R_{22} & R_{23} \\ R_{31} & R_{32} & R_{33} \end{bmatrix} \qquad (3.2)$$

Where, the Euler angles (in deg) are such that $\phi$ is equal to $90^o$-$\Lambda$, $\theta$ is the co-latitude or is equal to $90^o$ - $\lambda$ and $\Psi$ is equal to $\Psi$, i.e.,

$$R_{33} = Cos\theta = Sin\lambda$$

$$R_{32} = Sin\theta \cdot Cos\psi = Cos\lambda \cdot Cos\psi$$

$$R_{31} = Sin\theta \cdot Sin\psi = Cos\lambda \cdot Sin\psi$$

$$R_{23} = -Sin\theta \cdot Cos\phi = -Cos\lambda \cdot Sin\Lambda$$

$$R_{22} = -Sin\phi \cdot Sin\psi + Cos\phi \cdot Cos\theta \cdot Cos\psi$$

$$= -Sin\Lambda \cdot Sin\psi + Sin\Lambda \cdot Sin\lambda \cdot Cos\psi$$

$$R_{21} = -(Cos\phi \cdot Sin\psi + Sin\phi \cdot Cos\theta \cdot Cos\psi)$$
$$= -(Sin\Lambda \cdot Sin\psi + Cos\Lambda \cdot Sin\lambda \cdot Cos\psi)$$
$$R_{11} = Cos\phi \cdot Cos\psi - Sin\phi \cdot \cos\theta \cdot Sin\psi$$
$$= Sin\Lambda \cdot Cos\psi - Cos\Lambda \cdot Sin\lambda \cdot Sin\psi$$
$$R_{12} = Sin\phi \cdot Cos\psi + Sin\phi \cdot Cos\theta \cdot Sin\psi$$
$$= Cos\Lambda \cdot Cos\psi + Cos\Lambda \cdot Cos\lambda \cdot Sin\psi$$
$$R_{13} = Sin\theta \cdot Sin\psi = Cos\lambda \cdot Sin\psi$$

$$(3.3)$$

## 4-Tides on earth

In the terrestrial context, point "A" is the ocean surface at the port of interest or the laboratory and "B" the geocenter so ζ is the radius vector from the laboratory location to the geocenter. It should be noted that if the material of the earth is perfectly rigid then every acceleration observation, relative to the ground, automatically provides an (analog) differential measurement with respect to the earth's center of gravity.

Traditional calculations [57-63] of terrestrial tides is by series expansions of the scalar potential with harmonic functions. In the next sub-section our vectorial method will be described briefly.

### A- Direct vectorial computation

Here the tidal vector is computed directly from the expressions for the Newtonian gravitation [64] at the two locations. The position vector $R(t)_{cel}$ of the tide producing celestial body is defined as follows:

$$\vec{R}(t)_A \equiv \vec{R}(t)_{Topo-cel} \equiv \vec{R}(t)_{cel} \qquad (4A.1)$$

Where the Cartesian coordinates of $R(t)_{cel}$ in the topocentric coordinate system in the laboratory with x-axis along due east, y-axis, due north and z-axis is vertical (radially outward) pointing to the local zenith are given by,

$$\vec{R}(t)_{CEL} = \{X(t), Y(t), Z(t)\} \qquad (4A.2)$$

Let $r_e$, be the vector to the geocenter from the origin of the "topocentric" coordinate system (laboratory), then the base line vector (coordinates of the center of the earth) are given by,

$$\vec{\varsigma} = \{0, 0, -r_e\}. \qquad (4A.3)$$

The position vector of the celestial body relative to the geocenter, $R(t)_B$ is as follows

$$\vec{R}(t)_B = \vec{R}(t)_{Topo-cel} - \vec{\varsigma} \qquad (4A.4)$$

Where the Cartesian components of $R(t)_B$ in the lab frame are given by,

$$\vec{R}(t)_B = \{X(t), Y(t), Z(t) + r_e\} \qquad (4A.5)$$

The corresponding magnitudes are given by,

$$R(t)_A = \sqrt{X(t)^2 + Y(t)^2 + Z(t)^2}$$
$$R(t)_B = \sqrt{X(t)^2 + Y(t)^2 + \{Z(t) + r_e\}^2} \qquad (4A.6)$$

To avoid confusion by an over abundance of indices, in the following we will show only the essential subscripts. The correlation or difference in acceleration, $\Delta \boldsymbol{a}(\boldsymbol{R})_{cel}$ of the test particle in the laboratory relative to the geocenter is given by,

$$\Delta \bar{a}(t)_{cel} = GM[\frac{\bar{R}(t)_A}{\left|R(t)_A\right|^3} - \frac{\bar{R}(t)_B}{\left|R(t)_B\right|^3}]_{cel} \qquad (4A.7)$$

The x- component in the laboratory frame are given by

$$\Delta a_x(t)_{CEL} = GMX(t)\Phi(R_A, \varsigma, t)] \qquad (4A.8)$$

Where,

$$\Phi(R_A, \varsigma, t) = \frac{1}{\{X(t)^2 + Y(t)^2 + Z(t)^2\}^{3/2}} - \frac{1}{\{X(t)^2 + Y(t)^2 + (Z(t) + r_e)^2\}^{3/2}} \qquad (4A.9)$$

The y component of $\Delta \boldsymbol{a}$ is similarly obtained as:

$$\Delta a_y(t)_{CEL} = GMY(t)\Phi(R_A, \varsigma, t)] \qquad (4A.10)$$

The vertical or z component of $\Delta \boldsymbol{a}_z$ is a little more involved but is as follows,

$$\Delta a_z(t)_{cel} = GM[Z(t)\Phi(R_A, \varsigma, t) - \frac{r_e}{\{R(t)_B\}^3}] \qquad (4A.11)$$

We may also express $\Delta a(t)$ in horizontal-polar components $H(t)$, the declination $\theta(t)$ (direction between east and the net horizontal tide) and the vertical component $Z(t)$ as follows,

$$H(t) = (\Delta a_x(t)^2 + \Delta a_y(t)^2)^{1/2}$$

and

$$\theta(t) = \arctan[\frac{\Delta a_y(t)}{\Delta a_x(t)}]$$

The vertical component is as given by equation 4A.11 above.

## B- Sources of astronomical data

With the advent of the World Wide Web on the internet, astronomical and ephemeral information are readily available on the internet. For celestial vectors and coordinate data NASA's on-line-ephemeris at the JPL site [63] is very popular. The tidal potential catalogues [57-62] use JPL ephemeredes data. Another noteworthy web- resource is the Naval Observatory Vector Astrometry Subroutines, NOVAS [65]. NOVAS utilize matrices and vector formulation [65, 66], instead of traditional harmonic formulae of traditional spherical geometry so that the distances and coordinates of celestial objects are determined with greater rigor and higher precision.

**5-Corrections for earth's rotation**

Starting with Ehrenfest in the early 1900 rotation has been a challenging problem in Relativity. The resolution of the Ehrenfest paradox, understanding of the Sagnac effect and the Goedel Universe has not removed these difficulties. Still, today, many open questions remain. For example: what does it mean that the empty (non-flat) space-time rotates – does the axis of rotation gain the absolute status in the Newtonian sense? Why – standard notion of causality breaking Closed Time Curves (CTC) - are so abundant in solutions of the Einstein equations with rotational symmetry? These and many other questions persist and an extensive literature is available elsewhere [67-74]. Thus, here we will focus on the Newtonian description [45] only, namely fictitious forces are to be introduced to account for the non-inertial frames, so that

$$\vec{F}_{iner} = \vec{F}_{non-iner} + \vec{F}_{ficti} \qquad (5.1)$$

Because the astronomical forces and hence acceleration is small it is possible that fictitious forces can be significant and needs to be considered with care. The acceleration can be written as,

$$\vec{a}_{non-inert} = \frac{\vec{F}_{iner} - \vec{F}_{fict}}{m} \qquad (5.2)$$

or

$$\vec{a}_{non-iner} = \frac{\vec{F}}{m} + \left[ -\frac{d^2\vec{R}}{dt^2} + (\frac{d^2\theta}{dt^2}) \times \vec{r}_e + 2(\frac{d\theta}{dt}) \times (\frac{d\vec{r}}{dt}) + (\frac{d\theta}{dt}) \times (\frac{d\theta}{dt}) \times \vec{r}' \right] \qquad (5.3)$$

The terms in the square bracket arise from independent mechanisms and hence are often labeled as follows

$$a_{non-iner} = a_{inert} + acceleration(translational + azimuthal + coriolis + centrifugal) \qquad (5.4)$$

Hence, the acceleration as measured in the non-inertial frame is the acceleration caused by the Newtonian force (acting in an inertial frame) plus the vector sum of four additional components. Because the rotational velocity of the earth along the polar axis is $\omega \sim +72.9$ μRad/s, in terrestrial applications most deviations from inertial frames arise from the velocity dependant Coriolis contribution. In the laboratory if the object is free to undergo relative motion as with the Foucault pendulum then the Coriolis term can be significant. Furthermore, in a laboratory the mechanical acceleration of servo systems and mountings can give rise to the translational contribution, for measurements with accelerating turntables the azimuthal term may not be ignorable either. Finally, the centrifugal term should be included in calibrating for the local free fall acceleration (g).

So the corrected tidal acceleration of the test particle in the laboratory can now be written as follows,

$$a(t)_{in-cel} = GM_{cel} \cdot \left[ \frac{R(t)_{cel}}{|R(t)_{cel}|^3} - \frac{R(0,t)_{cel}}{|R(0,t)_{cel}|^3} \right] + a_{fict} \qquad (5.5)$$

## 6- Sensitivity and probable error estimation

The sensitivity ($\Delta$) of the tidal acceleration to the changes $\delta\alpha_i$ in the variables $\alpha_i$ can be easily determined as follows:

$$\Delta^2 = \sum_k (\frac{\partial \Delta \alpha_k}{\partial \alpha_k})^2 \cdot (\delta \alpha_k)^2$$

$$= \{ \sum_j^{N1} (m_j \cdot \frac{T_j}{R_j})^2 \} \cdot (\delta G)^2 + \sum_j^{N1} \{ G^2 \cdot (\frac{T_j}{R_j})^2 \cdot (\delta m_j)^2 + (a_j)^2 \cdot \sum_i^{N2} (\frac{\partial T_{ji}}{\partial \alpha_i})^2 \cdot (\delta \alpha_i)^2 \}$$

$$(6.1)$$

In equation 6.1, N1 is the number of celestial objects and N2 (N2>N1) the total number of variables in the tidal functions. Analysis of sensitivity is helpful in apparatus design and critical in the success of any experimental project. Equation 6.1 also expresses the probable error in a measurement of the acceleration signal in terms of the errors associated with the variables.

## 7- Applications of tidal accelerometry

In this section we will point out a number of implications and applications of tidal accelerometry. For specificity the focus will be on earth based accelerometric observations that is data obtained without the use of spacecrafts. First let us note that tides can introduce errors in a high precision experiment in a number of ways. Quite generally tidal contributions can be significant if the observations events are separated significantly in space –time. Even at the same location, as fig.3 shows just an hour's difference in time can cause measurable changes in the back ground to take place. For example in a G- experiment such steps may involve moving the field masses from "near" to "far" (or above and below) positions followed by adjustments and measurements then by this interval the tidal strengths can shift enough to produce errors. Shortly we will describe a case study of an experiment and show how the design of the apparatus with large spatial separation makes it particularly susceptible to tides. We will also indicate how these influences can be employed to calibrate the system and improve signal to noise ratio. However, with accurate astronomical inputs and data with sufficient statistics these tidal features can greatly increase the confidence level of the experiment.

After that we will consider a toy model and compute a set of synthetic tidal data with noise and extend Newton's idea about computing celestial mass from tidal observations. Another application is in locating non-radiating dark objects [75,76] and black holes.

A- Case study of a gravity experiment. Improving confidence level by de-trending of tidal signals.

Newton & Cavendish's universal constant of gravitations or the "Big-G" is the first constant of nature to be introduced in a physical theory however it is also amongst the least precisely known constants [77-83]. For example the most recent "2006 CODATA recommended value" [84] which became available in March 2007 is 6.674 28 x $10^{-11}$ $m^3$ $kg^{-1}$ $s^{-2}$ Relative standard uncertainty (RSU) 1.0 x $10^{-4}$ . In comparison the RSU in the Planck constant h is 5.0 x $10^{-8}$ . Decreasing the uncertainty in G has become a "Cause Celebre".

Not surprisingly, determination of G requires the sensitive measurements of the gravitational acceleration of a "test mass" as the position of a "field mass" is varied in a prescribed manner. To avoid sensitivity to material property variations in the torsion fiber, many modern G-experiments employ torsion strips or multi-fiber systems. This design couples the angular (rotational) degree of freedom with the vertical direction hence results in a compromise, trading uncertainties in the elastic energy with that in gravitational potential.

One class of designs [85-87] in gravitational experiments suffers from two opposing requirements: the needs to reduce (i) the overall size of the apparatus and (ii) cross influence between test and the control arms (counter weight) - hence they are physically close in the (say) horizontal direction but kept far apart along the vertical axis.

This design conflict is endemic and has afflicted many Big-G and gravity experiments since C.V. Boys pointed out on basically dimensional arguments, because the stiffness constants of the force measuring element (sic, torsion fiber) scales at a higher power with size (typically $\sim d^4$) than the load (weight $\sim$ volume $\sim d^3$) and load bearing capacity (breaking load $\sim$ cross sectional area $\sim d^2$) so higher force sensitivity is achieved when the apparatus is scaled down to the smallest possible size. Boys himself was aware of this conflict hence to reduce "cross torque" on his torsion pendulum he was forced to suspend the two tests masses at two different heights from the pendulum beam. However, as indicated earlier increased space-time separation can be detrimental and Boys failed to realize that his design increased the sensitivity to astronomical tides.

Next we outline of how tidal influences can be magnified by apparatus design, used for calibration and how it can be - de-trended. For specificity consider a G-experiment [83] and the use of a beam balance. The acceleration data for nine days as obtained from published figures are given in table-II and are shown in figure 6 (red dots). Data plotted as a graph shows a systematic drift with time of observation; from a linear curve fit (solid red line) one obtains a slope of -.3986 $nm/s^2$ per day. This slope quantitatively measures the rate of decreases in the mean value of the acceleration.  In the same figure the values of the de-trended data (blue squares) are also plotted. A de-trending is performed by numerically removing for the drift contribution. B After this corrective process the mean drift rate is reduced by two orders of magnitude (solid blue line). However, notice that this de-trended, "drift-free" data contains three pronounced undulations.

A harmonic analysis (green curve) of the drift-free behavior is shown in fig. 7. The fit provides a harmonic amplitude of 15.8 $nm/s^2$, time period of about 27 (+/-1) days and a correlation coefficient R of 0.8; considering that the nine data points (table–II) are obtained from a printed figure with graphical precision the correlation is quite adequate. Furthermore two of these peaks correlate with the days of full moon (August, 4th, &

September 8th, 2001) and the minimum with the new moon on August 19th. Note that on August 19 (new moon) the moon was in perigee.

When this harmonic contribution is removed or subtracted out of the data then one   obtains the tide corrected values of acceleration that is free of both linear drift and harmonic modulations. In Table II all three sets of data are shown. The raw acceleration data (red dots) is compared with the corresponding corrected values (green squares) as described, the resulting improvement on the scatter is clearly as shown in figure. 8.

Hence the scatter between maximum and minimum values (784.88 & 784.84) in the raw observations [83] is reduced in the de-trended set down to(784.87 & 784.85) from about 40 ppm to 2 ppm, likewise the standard deviation decreases from 0.015 down to 0.0008 after removing the systematic extraneous signals. This process can further be improved by de-trending for the diurnal tidal effects if the experimental results obtained at different times of the days are also available.

The Doodson type harmonic analysis [88-90] (fig.7) was chosen because tidal calculations, either by potentials or direct vectorial methods (sec. 4 A &B) would require further information (not available in the publications) regarding the exact geographical coordinates of the place and time of observation. Nevertheless, the fact that the quantitative estimates of the strength and time period plus that the reported data points so accurately match the phases of the moon is a strong evidence for tide.

When the tidal vector components for all the observations times are calculated then the diurnal, semidiurnal and the lunar month modulation amplitudes all can provide independent standards of forces as well as time. Indeed if one is carefully measuring forces or accelerations then making sure that verifiable tidal signals are indeed detected, provides a good calibration plus additional credibility.

Often it is stated that "… a beam balance is inherently insensitive to tidal forces…". However, it is not so for all apparatus design that employs a beam balance! All depends on the symmetry or asymmetry in the mass and "counter mass" arrangement. For examples the problem is acute in configurations where one mass is hung at a large separation (~2 m) away from the beam [83] whereas the counter weight is "in-line" with the beam. As shown in fig. 9 due to this large difference in the lever arm distances, such a lop-sided balance design can create a huge mechanical advantage.  Such a design with large vertical separation greatly exaggerates and amplifies the tidal torques especially that associated with the horizontal component of the tidal force.  In a high precision experiment, it is important to understand such trade-offs and make appropriate modifications in hardware and procedural designs.

B-Determination of celestial masses without Kepler's law

As was mentioned earlier, Isaac Newton's attempts to compute lunar mass ($M_m$) is an example finding inertia from earth based water tide observations. Briefly, Newton was motivated by the observations (i) the highly successful "1-2-3" law of Kepler is inapplicable for $M_m$ because moon has no natural satellites. (ii) At that time tides were the only available measure of moon's gravitational force. To find $M_m$ Newton used his estimate of the sun-earth masse ratio and maximum and minimum tidal changes in the deep ocean water heights; he computed the moon-earth mass ratio to be 1/39.79 about twice the presently accepted value. However, as will be outlined next, the tidal method can be extended beyond Newton's scheme for lunar mass and compute the values of solar system masses and in principle any sufficiently massive celestial object from tidal observations. Alternate methods of determining solar system masses is of interest because of cross checking as well as in increasing the reliability of ephemerides used in space craft guidance [91].

Here we will be considering the contributions from the n celestial objects with the unknown masses ($m_i$, i = 2,3,  n) so equation for $\Delta a(t)$ tidal acceleration signal [section 4.B] is re-written as

$$\Delta a(t) = [\{m\Theta(R;t)\}_1 + ... + \{m\Theta(R;t)\}_i + ... + \{m\Theta(R,t)\}_n]$$
(7B.1a)

For the sake of specificity say that a set of N values of $\Delta a(t)$ are available. Like most astrophysical experiments typically there is an overabundance of observations or data points so the analysis requires special considerations [92,93] and equation 7B.1a constitutes an over-determined set of relationships between the n unknown masses and N data points. Hence this set offers multiple solution strategies, here we describe a straight-forward method namely solution by the linear least square (LLSS) scheme [94].

To consider the general problem let us normalize equation 7B.1 by dividing out all the n+1 terms in the LHS and RHS by $\Theta(R,t)_n$ to obtain:

$$C_1^1 m_1 + ... + C_1^j m_j + ... + C_1^{n-1} m_{n-1} + m_n = D_1$$

$$C_i^1 m_1 + ... + C_i^j m_j + ... + C_i^{n-1} m_{n-1} + m_n = D_i$$

$$C_N^1 m_1 + ... + C_N^j m_j + ... + C_N^{n-1} m_{n-1} + m_n = D_N$$
(7B.1b)

In the linear least square scheme (LLSS) the over determined system of simultaneous equations (7.B1b) comprising of N data points ($D_j$) for the n unknowns $m_j$'s is represented by the three matrices as follows:
(i) the (nxN) coefficient matrix ($C$),defined as

$$C = \begin{bmatrix} 1 & C_1^1 & .. & C_1^{n-1} \\ 1 & .. & .. & C_2^{n-1} \\ 1 & C_j^1 & .. & C_j^{n-1} \\ 1 & .. & .. & C_N^{n-1} \end{bmatrix}$$

(ii) the (1xN) data matrix ($D$), defined as

$$D = \begin{bmatrix} D_1 \\ D_j \\ D_N \end{bmatrix}$$

and (iii) the (1x n) solution or mass matrix, ($M$),

$$M = \begin{bmatrix} m_1 \\ m_j \\ m_n \end{bmatrix} \tag{7B.1c}$$

Where for brevity only a few of the elements of the matrices including the generic jth term are shown in equation 7B1c. The strategy of LLSS is to construct a suitable nxn matrix ($H$) and solve for M by a matrix inversion process, as follows:

$$H = C^T C$$

or,

$$H = \begin{bmatrix} N & \sum C_i^1 & .. & \sum C_i^{n-1} \\ .. & \sum (C_i^1)^2 & \sum C & .. \\ \sum C_i^j & .. & \sum (C_i^j)^2 & .. \\ \sum C_i^{n-1} & .. & \sum C_i^{n-2} C_i^{n-1} & \sum (C_i^{n-1})^2 \end{bmatrix}$$

and,

$$C^T D = \begin{bmatrix} \sum D_i \\ \sum C_i^1 D_i \\ \sum C_i^j D_i \\ \sum C_i^{n-1} D_i \end{bmatrix}$$

So that

$$M = H^{-1} C^T D \tag{7B.1d}$$

Notice that because there n, $\Theta(R,t)_n$ coefficients there n normalization choices to reduce 7B1a to 7B1b, there are n solution matrices ($M$) providing a range of possible values.

To make the discussion concrete and visual let us consider the determination of two unknown masses say $m_1$ & $m_2$. In this case the LLSS procedure can be visualized as a least square straight line fit to the scatter plot of Y (D) vs X (C ) and the two unknowns are obtained as the intercept and slope. There are two ways of defining these plotting variables depending on the two choices of normalizations, namely $\Theta(t)_1$ or $\Theta(t)_2$ . Normalization with respect to $\Theta(t)_1$ reduces equation 7B.1 to

$$\frac{\Delta a(t)}{\Theta(t)_1} = m_1 + m_2 \frac{\Theta(t)_2}{\Theta(t)_1} \tag{7B.2a}$$

or

$$Y(t) = m_1 + m_2 X(t) \qquad\qquad (7B.2b)$$

Where, $Y(t) = \Delta a(t)/\Theta(t)_1$ and $X(t) = \Theta(t)_2/\Theta(t)_1$. In this case a linear fit of the scatter plot of Y vs X data will have $m_1$ as the intercept and $m_2$ as slope. (ii) In the second option normalization is performed with respect to $\Theta(t)_2$ and equation 7B.1 reduces to

$$\frac{\Delta a(t)}{\Theta(t)_2} = m_2 + m_1 \frac{\Theta(t)_2}{\Theta(t)_1} \qquad\qquad (7B.3a)$$

or

$$Y(t) = m_2 + m_1 X(t) \qquad\qquad (7B.3b)$$

Where, $Y(t) = \Delta a(t)/\Theta(t)_2$ and $X(t) = \Theta(t)_1/\Theta(t)_2$. Now a linear fit of the scatter plot of Y vs X data will have $m_2$ as the intercept and $m_1$ as slope. This leads to two distinct ways to compute the desired masses and in principle improves the reliability of the computation.

To illustrate the LLSS technique we perform a toy model calculation of 72000 observations. The observed signals are considered to be due to two sources $m_1$ & $m_2$ plus some white, random noise from a couple different sources. We consider a simple situation where the laboratory is on the surface of a planet and the two celestial masses move at slightly different rates in the equatorial plane of the planet. Let $m_1$ be two thousand times that of $m_2$, i.e., $m_2 = 1$ (R.U) and $m_1 = 2000$ (R.U), but is more distant such that $\Theta(t)_1/\Theta(t)_2$ is say $\sim 1 \times 10^{-4}$ over the time period in question. The theoretical values of $\Theta(t)$ for both objects at each of the 7200 observation times of were also tabulated. One per cent each of two random numbers $Rnd_1$ & $Rnd_2$ ($-1 < Rnd < +1$, $Rnd_{ave} = 0.0$) were added to the calculated tides and create the simulated experimental data $\Delta a(t)$. The $Y(t)$ and $X(t)$ are calculated following the first normalization procedure described above, prior to division by $\Theta_1$, care was taken to exclude the points with zero or nearly zero values of $\Theta_1$. The tidal signal and analysis by the first normalization are shown in fig. 10, the LLSS values are $m_1 = 1992$ (compared with 2000 exact) and $m_2 = 1.0012$ (1.000 exact). In the second normalization process LLSS determined $m_1$ & $m_2$ to be 2021 and 0.99794 respectively.

## C- Other applications and conclusions

The results of tidal calculations shown earlier contain both inertial and spatial information hence all of them can be obtained by the inversion of observations with adequate resolution. So that observations of tidal acceleration can be utilized to locate geographical location and astronomical time. With sufficient sensitivity it can be possible to study the uniformity of lunar mass distribution without space crafts, but by tidal signals from detectors based on the earth. Furthermore absent corrections the tidal signals can poison other high precision data. Perhaps the situation is similar to the inclusion of relativistic corrections in the satellite based geo-position system (GPS) [95-97]; that is to say, precision force and acceleration data must account for tidal contributions. Tidal signals can also be complimentary to GPS. Additionally, inspired by the metrological advances in pulsar timing array (PTA) and laser interferometric, gravity wave detector [98-99] technologies we reason that with sufficient development the proposed new technique can be important in evaluating some of the contending theories in cosmology.

Newtonian gravitation, **g** is non-radiative, infinite range and all pervasive. Spatial correlation of g gives rise to tides which can be computed as a vector difference. Tidal signals contain quantitative astrophysical information and affect high precision measurements. Guided by modern advances in gravity wave detectors we argue that it is important to develop high precision accelerometry. With a resolution of $\Delta R \sim 10^{-9}$ m it will be possible to determine solar system masses and detect the SMBH at the center of our galaxy. Observations of the gravitational correlation can potentially open up yet to be explored vistas of the cosmos.

## 8- Acknowledgements

This article grew from our "talking point" notes initially prepared for informal presentations, colloquium and discussions. We wish to thank Profs Barry Barish for sharing his experience especially discussions of tidal modulations[99] in the LIGO drift velocity and Donald York for conversations on challenges in contemporary Astrometry.

No acknowledgement on a work on gravitation by this group can be complete without respects to the memories of camaraderie and professional rivalry of the late Prof. Jeeva Anandan. He passed away in another hot and humid, Columbia summer (2003) after a valiant battle against ALS but never lost his dignity and continues to inspire at least two amongst us (TD & PM). Also thanks to Prof. Pawel mazur, the remaining member of the famed "Foundations" troika at USC –Aharonov, Anadan and Mazur. This work was partially supported by the National Science Foundation (Awards: OCE-0451989 and OCE-0535893).

**Table –I**

| Celestial Object | Maximal tidal acceleration* |
|---|---|
| Moon | $1.11\ \mu m/s^2$ |
| Sun | $0.51\ \mu m/s^2$ |
| Venus | $0.06\ nm/s^2$ |
| Jupiter | $8.0\ pm/s^2$ |
| Mars | $3.0\ pm/s^2$ |
| Saturn | $3\times10^{-13}\ m/s^2$ |
| SMBH (@ 10 kpc) | $10^{-24}\ m/s^2$ |
| Ten Solar mass (@ 1 kpc) | $10^{-26}\ m/s^2$ |
| Ten Solar mass (@ 10 pc) | $10^{-20}\ m/s^2$ |

\* Here we use SI unit, Galileo or Gal is short for another accepted unit of acceleration, $1 Gal = 1 cm/s^2$ that is also common in the accelerometry literature.

**Table -II**

| Day | Raw accel ($\mu$-m/s$^2$) | Lin-De-Trend ($\mu$-m/s$^2$) | Cor-accel ($\mu$-m/s$^2$) |
|---|---|---|---|
| 212.00 | 784.87 | 784.86 | 784.85 |
| 219.00 | 784.89 | 784.89 | 784.87 |
| 223.00 | 784.88 | 784.88 | 784.88 |
| 228.00 | 784.85 | 784.85 | 784.87 |
| 232.00 | 784.86 | 784.86 | 784.88 |
| 238.00 | 784.87 | 784.87 | 784.87 |
| 243.00 | 784.89 | 784.89 | 784.88 |
| 247.00 | 784.86 | 784.87 | 784.86 |
| 252.00 | 784.84 | 784.86 | 784.86 |

**Table-III**

| | Raw accel | Lin D-Trend ($\mu$-m/s$^2$) | Corrected accel. |
|---|---|---|---|
| No of pts | 9 | 9 | 9 |
| Sum | 7063.8076 | 7063.83 | 7063.8112 |
| RMS | 784.86751 | 784.87 | 784.86791 |
| Std Deviation | 0.015024535 | 0.014039295 | 0.0084401811 |
| Variance | 0.00022573666 | 0.0001971018 | 7.1236657e-05 |
| Std Error | 0.0050081784 | 0.0046797649 | 0.0028133937 |

**References**

1. To the best of current information it seems that compared with the invention of the microscope the historical facts relating to the invention of the telescope are obscure. The history of lenses is not fully available; but, from hieroglyphical records it appears that Egyptians as early as the 5th century BC were familiar with the properties of simple lenses. Emperor Nero is reported to have used gem stones for corrective purposes. But the record of wear able contraptions as eye glasses (near sighted) and spectacles (far sighted) starts in Italy some time in the 13th century. It took about two more centuries for telescopes to appear. Also see Controversy over telescope origin". BBC News. 16 September 2008. Retrieved 2009-07-06

2. There is evidence that by the 1570's scholars in Britain such as John Dee and Thomas Diggs have constructed working telescopes.  Also one Thomas Harriot who was an Oxford graduate (1577), an algebra scholar, a navigator, cartographer, and an explorer of Virginia under Sir Walter Raleigh. On July 16, 1609 Harriot documented astronomical observations of the moon's surface (at 6X) and December 1610 of sunspots; thereby preceding Galileo's astronomical observations by several months.

3. Hans Lippershey's demonstration of a telescope, and sighting of the starts was in 1608. It is interesting to note that although Lippershey is credited to have been the first to publicize the design of the apparatus but his patent application failed because the States-General of the Netherlands observed that "others have also claimed the same invention"!

4. Galileo's ambidexterity in science as well as in publicity was the single most important factor in the popularity and success of telescopes. Even the name "telescope" was coined in 1611, by one of his "patrons", prince Frederick Sesi at a reception for a demonstration by Galileo of his apparatus; http://galileo.rice.edu/sci/instruments/telescope.html . Arguably, it is imperative that the scientist realizes the importance of his/her work and is obligated to do what it takes for the community to notice this importance. Please also see Michael Berry, Aptly named Aharonov-Bohm effect has classical analogue, long history, Phys. Today, **63**, 8(2010)

5.  Michael S. Turner, The back page, APS News, **18**, March 2009.

6. D. Nicholson et al, Results of the first coincident observations by two laser-interferometric gravitational wave detectors, Phys. Letts. A, 218 175- 180 (1996)

7. Kip S. Thorne, Probing black holes and relativistic stars with gravitational waves, arXiv:gr-qc/9706079v1 26 Jun 1997B.

8. Abbott et al, Einstein@Home search for periodic gravitational waves in LIGO S4 data, Phys. Rev. D **79**, 022001  [29 pages] (2009) doi: 10.1103/PhysRevD.79.022001; The LIGO Scientific Collaboration & The Virgo Collaboration, An upper limit on the stochastic gravitational-wave background of cosmological origin, doi:10.1038/nature08278. Also, doi:10.1038/460964a

9. Geoff Brumfiel, General Relativity, to catch a wave, **Nature, 417**, 482-484 | doi:10.1038/417482aLIGO ( 2002)

10. Geoff Brumfiel, Gravity-wave hunt stalled, Nature doi:10.1038/news.2008.1105 (2008)

11. Jeffrey J. Love, Magnetic monitoring of earth and space, Physics Today, **61**, 31-37 (2008). See also Phys Today 16 (2001).

12. http://www.intermagnet.org

13. In this article we concentrate on the gravitational fields that are of extra-terrestrial origin and the implications are different from what was discussed by Timir Datta, Gravitational gauss law in the laboratory: A 21 century Archimedes problem, 2006physics...2163D

14. Elizabeth S. Cochran, John E. Vidale, Sachiko Tanaka, Earth Tides can Trigger Shallow Thrust Fault Earth Quakes, Science 306, 1164-1166 (2004) doi:10.1126/Science.1103961

15. Albert Einstein, *The meaning of Relativity*, 5th ed. Princeton, (1921).

16. W. Pauli, *Theory of reltivity*, Dover, New York (1981).

17. T. Datta & M. Yin , Do Quantum Systems Break the Equivalence Principle, arXiv:0908.3885v1

18. Please also see http://www.technologyreview.com/blog/arxiv/24050/

19. R. Hanbury Brown and R. Q. Twiss (1956). "A Test of a New Type of Stellar Interferometer on Sirius". *Nature* **178**: 1046–1048. doi:10.1038/1781046a0. Also see- http://en.wikipedia.org/wiki/Hanbury_Brown_and_Twiss_effect

20. T. Datta, Ming Yin, Andreea Dimofte, M. Bleiweiss, A. Lungu, Jafar Amirzadeh, and W. E. Sharp, Tidal Effects on Gravity Experiments with a Balance, Phys. Rev. Lett. 91, 109001, (2003): 10.1103/PhysRevLett.91.109001

21. Joshua P. Schwartz et al. A free –fall determination of the Newtonian constant of gravity, Science, 282,2230 (1998) doi:10.1126/science.282.5397.2230

22. Achim Peters, Keng Yeow Chung & Steven Chu, Measurement of gravitational acceleration by dropping atoms, Nature, **400**, 849 (1999); also, Holger Muller et al., Nature,**463** (2010) doi: 10.1038/nature08776

23. L. Arnaudon et al, Accurate determination of the LEP beam energy by resonant depolarization, Zeitschrift Fur Physik C, Particles and fields, **66**, 45-62(1995)

24. L. Arnaudon et al, Effects oftidal forceson the beam energy in LEP, IEEE, PAC 1993, 44-46 (1993)

25. Isaac Newton, *Philosophiae naturalis principia mathematica*, Streater, London (1687).

26. *Newton's Principia*, translated by Andrew Motte, (1729).

27. Please see the on-line version of the 1$^{st}$ American edition of Andrew Motte's translation - http://rack1.ul.cs.cmu.edu/is/newton/

28. S.Chandrasekhar, *Newton's Principia for the Common Reader*, Clarendon Press,Oxford (1995)

29. C.W. Meissner, K. S. Thorne , J. A. Wheeler, Gravitation, Freeman, New York, (1973).

30. O.Schnurr, A.F.J. Moffat, A. Villar-Sbaffi, N. St-Louis & N.I. Morrell, A first-orbital solution for the very massive 30 Dor main-sequence WN6h+O binary R145, *Mon. Not. Roy. Astron. Soc*. **000**, 1-15 (2008)

31. Leon Blitzer, Morris Weisfeld, Albert D. Wheelon, Perturbations of a satellite's orbit due to the earth's oblateness, **Jr. Apld. Phys**, **27**, 1141-1149 (1956)

32. L.M. Celnikier, Weighing the earth with a sextant, **Am. J. Phys**. **51** , 1018-1020 (1983)

33. W.J. Gallagher, Mass of the moon from satellite observations, *Science*, **128**, 1207 (1958)

34. H.L. Armstrong, Weighing the moon, Am. J. Phys. **32** , 985-986 (1966).

35. J.D. Anderson, G.E. Pease, L. Efron, R.C. Tausworthe, Celestial Mechanics Experiment, *Science,* **158**: pp 1689-1690 (1967) doi: 10.1126/science.158.3809.1689

36. S. K. Wong & S. J. Reinbold, Earth-Moon Mass Ratio from Mariner 9 Radio Tracking Data**,** *Nature* **241**, 111 - 112 (1973) doi:10.1038/241111a0.

37. V.P. Dolgachev, Long-period perturbations of the orbital elements of artificial earth satellites due to lunar gravitation, Moskovskii Universitet, Vestnik, Seriia III-Fizika, **Astronomia, 15**, 591-596 (1974)

38. P.J. Shelus, R.L. Ricklefs, J.R. Wiant, J.G. Ries, Lunar and artificial satellite laser ranging: The use of queue scheduling and worth functions to maximize scientific results, *Bull. Am. Astro. Soc*, **35**, 1038 (2003)

39. C.C. Goad & B. C. Douglas, Tidal acceleration of the moon deduced from observations of artificial satellites, *Nature*, **266**, 514-515 (1977)

40. Stephen B. Fels, J. T. Schofield & David Crisp, Observations and theory of the solar semidiurnal tide in the mesosphere of Venus, *Nature*, **312**, 431-434 (1984)

41. *nssdc.gsfc.nasa.gov/planetary/factsheet/**moon**fact.html*

42. August, 2009, IAU current best estimate (CBE) as determined by the NSFA WG Mm/Me= 1.230 003 71 e-2  http://maia.usno.navy.mil/NSFA/CBE.html

43. N. Kollerstrom, "Newton's Two Moon-Tests", British Journal for history of Science, **24**, pp 369-72 (1991).

44. Pierre-Simon de Laplace, Trait´e de M´ecanique C´eleste, Tom IV, Par 2, Courcier, Paris. (1805).

45. Jean le Rond d'Alembert, Recherches du Systeme du Monde, Paris (1754-56)

46. Please see discussion regarding Newton's Lunar mass problem by C. Wilson in: 'Perturbations and Solar Tables from Lacaille to Delambre', Archive for History of Exact Sciences, **22**, pp 53-188, 88 (1980).

47. C. Wilson, D'Alembert versus Euler on the Precession of the Equinoxes , Archive for History of Exact Sciences **37**, 233-273, 242 (1987).

48. R. S. Westfall, 'Newton and the fudge factor', Science, **179**, pp. 751-8 (1973).

49. N. Kollerstrom, "Newton's Lunar Mass Error", Jrnl of the British Astronomical Association, **95**, pp. 151-153, (1985), and http://www.dioi.org/kn/newtonmoonerror.htm

50. Dave Waltham, Anthropic Selection for the Moon's Mass, Astrobiology, **4**, pp 460-468 (2004), doi:10.1089/ast.2004.4.460.

51. Alan P. Boss, The Origin of the Moon, *Science*, **231**, pp. 341 − 345 ( 1986) doi : 10.1126/science.231.4736.341

52. Carsten Münker, Jörg A. Pfänder, Stefan Weyer, Anette Büchl,Thorsten Kleine, Klaus Mezger, Evolution of Planetary Cores and the Earth-Moon System from Nb/Ta Systematics, *Science,* **301**, pp. 84 − 874 (2003): doi: 10.1126/science.1084662

53. James M. D. Day, D. Graham Pearson, Lawrence A. Taylor, Highly Siderophile Element Constraints on Accretion and Differentiation of the Earth-Moon System, *Science,* **315**, pp. 217 − 219 ( 2007) doi: 10.1126/science.1133355

54. Arnold Sommerfeld, *Mechanics*, (Lib. Of Cong. Cat # 50-8749) Academic press, NY

55. L.D. Landau and E.M. Lifshitz, *Mechanics*, (ISBN 0 7506 2896 0) Butterworth-Heinemann, Oxford (1976)

56. H. Goldstein, C.P. Poole, J. Safko, Classical Mechanics, 3$^{rd}$ ed. Addison Wesley, ISBN 0201657023, (2002).

57. G. Godwin, The Analysis of Tides, Univ. Toronto Press, (1972)

58. Rich Pawlowicz, Bob Beardsley, Steve lentz, Classical tidal harmonic analysis including error estimates in MATLAB using T_TIDE, Computers & Geosciences, **28**, 929-937 (2002)

59. Kevin Hamilton, Steven C. Ryan, Wataru Ohfuchi, Topographic effects on the solar semidiurnal surface tide simulated in a very fine resolution general circulation model**, J. of** Geophys Res**, 113**, D17144 (2008)

60. Hartmann, T. and H.-G. Wenzel, Catalogue of the earth tide generating potential due to the planets. Geophysical Research Letters, Vol.21, pp. 1991-1993 (1994)

61. Hartmann, T. and H.-G. Wenzel, Catalogue of the earth tide generating potential due to the planets. Bulletin d'Informations Marees Terrestres, Vol. 119, 8847-8880, Bruxelles (1994).

62. Hartmann, T. and H.-G. Wenzel, The HW95 tidal potential catalogue. Geophysical Research Letters, Vol. 22(24), pp. 3553-3556 (1995).

63. http://ssd.jpl.nasa.gov/?ephemerides

64. Gambrell, James; Wescott, Michael; Yin, Ming; Overcash, Dan; Voulgaris, George; Datta, Timir, A Study of Tidal Acceleration, Bulletin American

Physical Society, 76th Annual Meeting of the Southeastern Section of APS, abstract #PB.009 (November 11-13, 2009).

65. http://www.usno.navy.mil/USNO/astronomical-applications/software-products/novas/novas-c.

66. Kaplan et al Astron J. 97, 1197 (1989).

67. M Mason, Note on the retarded potential. Physical Rev, **15**,.312-316 (1920)

68. A. Jaques and J.S. Morgan, Rotation and Relativity Nature 118, 194-194 (07 August 1926) | doi:10.1038/118194d0

69. Nathan Rosen, Notes on rotation and rigid bodies in relativity theory, Phys Rev, 71, 54 (`1947)

70. Jeffrey M. Cohen & William J. Sarill, Centrifugal force and General relativity, *Nature* **228**, 849 (28 November 1970); doi:10.1038/228849a0

71. Marek A. Abramowicz, Centrifugal force: a few surprises, RAS monthly notices, **245**, 733 (1990)

72. Michael Weiss, Can you see the Lorentz-Fitzgerald contract? Or: Penrose-Terrel rotation
http://www.phys.ncku.edu.tw/mirrors/physicsfaq/Relativity/SR/penrose.html

73. G. Z. Machabeli, I. S. Nanobashvili and A. D. Rogava, Centrifugal acceleration surprises, Radiophysics and Quantum Electronics 39, Number 1, 26-30, DOI: 10.1007/BF02121461

74. Pawel Morawiec, Physical & geometrical aspects of de Sitter interior of a gravastar, PhD dissertation, USC Phys & Astronomy, August, 2010.

75. Massey R Rodes J; Ellis, R; Scoville, N; Leuthaud, A; Finoguenov, A; Capak, P; Bacon, D *et al*. "Dark matter maps reveal cosmic scaffolding". *Nature* **445** (7125): 286–290. (2007). doi:10.1038/nature05497

76. Bertone, Gianfranco, Hooper, Dan and Silk, Joseph, Particle dark matter: evidence, candidates and constraints, *Physics Reports*, **405**, 279-390 (2005).

77. Yin, Ming; Wescott, Michael; Overcash, Dan; Voulgaris, George; Cokkinides, George; Morawiec, Pawel; Datta, Timir, A new technique to determine the value of G, American Physical Society, 76th Annual Meeting of the Southeastern Section of APS, November 11-13, 2009, abstract #PB.008

78. Eugenie Samuel Reich, G-whizzes disagree over gravity, 1030, News, Nature,**466**, 26 August (2010).

79. Jun Luo, et al. Determination of The Newtonian Gravitational Constant G with Time of Swing Method, *Phys. Rev. Lett.* **102**, 240801 (2009).

80. H. V. Parks & J. E. Faller, A simple Pendulum Determination of the Gravitational Constant  http://xxx.lanl.gov/arXiv:1008.3203
To appear in *Phys. Rev. Lett.*

81. Riley Newman, Convergence (?) of G measurements- Mysteries Remain, http://www.phys.lsu.edu/mog/mog21/node12.html

82. CC Speake, Newton's constant and the 21$^{st}$ century laboratory.
http://rsta.royalsocietypublishing.org/content/363/1834/2265.full

83. W. Kündig, et al DOI: 10.1103/PhysRevLett.**91**.109002 ; Phys. Rev. Lett. **89**, 161102 (2002);also:10.1103/PhysRevD.74.082001

84. P.J. Mohr, B.N. Taylor, D. B. Newell, CODATA recommended values of the fundamental physical constants 2006, Rev of Mod Phys, **80**,633-730 (2008) http://physics.nist.gov/cgi-bin/cuu/Value?bg|search_for=G

85. C.V. Boys, On the Cavendish experiment, Proc. Roy. Soc., **46**,253-268 (1889);

86. C.V. Boys, On the Newtonian constant of gravitation, Philos. Trans. Roy. Soc., A **186**, 1-72 (1898).

87. Y.T. Chen & Alan Cook, Gravitational experiments in the laboratory, Cambridge Univ. Press (1993).

88. D E Cartwright, Tides: a scientific history, Cambridge University Press 2001

89. A.T. Doodson The harmonic development of tide-generating potential, Int. Hydrodynamic Rev, **31**, 37-61 (1954)

90. J. Proudman, Bio. Mem. Of Fellows of the Roy. Soc, **14**, 189-205 (1968)

91. D.J. Champion et al. "Measuring the mass of Solar-System planets using pulsar timing", *Astrophysical Journal Letts., ***720**:L201-L205 (2010). doi:10.1088/2041-8205/720/2/L201

92. Yeuncheol Jeong, Deviation of CMBR from a perfect blackbody caused by non-equilibrium radiation of fractal dust grains, PhD dissertation, Astronomy dept., UCLA (1994).

93. Popper, D. M., & Jeong, Y. C. Publication of the Astronomical Society of the Pacific (PASP), **106**, 184 (1994)

94. Howard Anton & Chris Rorres, Elementary Linear Algebra, 9[th] ed, John Wiley (2009).

95. Neil Ashby, Relativity and GPS. Physics Today, May ( 2002).

96. See, http://www.losangeles.af.mil/library/factsheets/factsheet.asp?id=5311.

97. Richard W Pogge, "Real-World Relativity: The GPS Navigation System"

98. Paul Demorest, Joseph Lazio & Andrea Lommen, Gravitational-wave detection via radio-pulsar timing, Physics Today, **63**, January - QUICK STUDY, (2010)

99. F. Rab & M. Fine, The effects of Earth tides on LIGO interferometers, LIGO-T970059-1-D (1997)

# Figure Captions

**Figure 1:** Behaviors of two nearby geodesics. In source free, empty or flat space (left) curvature (R) is zero the relative separation between two apples following their respective geodesics remains the same. The separation changes and geodesics are no longer parallel in curved space-time. Spatial differences in the geodesics reveal curvature and betrays the presence of a nearby object (right)

**Figure 2:** The changes in the weight with time of an object in our laboratory and correlation with the tidal water height at the port of Charleston, SC. The green rectangle highlights a period of the rapid changes in balance reading with time.

**Figure 3:** Examples of tidal influences in the laboratory; on the vertical free fall acceleration of a macroscopic test mass (bottom- shaded region), an absolute gravimeter (solid blue curve) and that of Cesium atoms in an atomic fountain (top- solid red curve); after Schwartz et al and Peters et al.

**Figure 4:** The LEP beam energy dependence on tidal variations. The data (solid red and blue dots) points are of graphical accuracy, an estimated ~ 1ppm ($\Delta E/E$) decrease in beam energy per μ-gal increase in g is shown by the diagonal straight line (green); after Arnaudon et al.

**Figure 5:** Coordinate transformation, (a) $P$ is the location ($\lambda$, $\Lambda$) of the laboratory on the surface of the earth in ***XYZ*** a geocentric coordinate. (b) Radial translation by ***r***$_e$ from the geocentric origin to surface centric origin is shown.

**Figure 6:** Acceleration data for nine days (red dots) obtained over a period of a month (after Kundig et al). The LSS fit shows a systematic drift with time of observation with a slope of ~ -0.3986 n-m/s$^2$ per day.

**Figure 7:** Harmonic analysis (green curve) of the drift-free acceleration (blue dots). The Doodson type fit provides harmonic amplitude of 15.8 nm/s$^2$, time period of about 27 (+/-1) days. Furthermore, the peaks correlate with the days of full moon (August, 4th, & September 8th, 2001); the minimum on new moon, is August 19[th] when moon was in perigee, all in excellent agreements with tidal behavior.

**Figure 8:** Improvement of confidence level by correcting for both harmonic changes and drift (green squares) hence the reduction of the scatter in the raw acceleration data (red dots). As described in the text after removing the systematic extraneous signals, the scatter between maximum and minimum values in the raw observations is reduced from ~ 40 ppm to 2 ppm, likewise the standard deviation dropped from 0.015 to 0.0008.

**Figure 9:** The design of a lop-sided balance creates a huge mechanical advantage. A large vertical separation greatly exaggerates and amplifies the tidal torques especially that associated with the horizontal component of the force.

**Figure 10:** A synthetic tidal signal due to two masses plus noise. Inset shows the scatter plot and linear least square analysis for the two unknowns, $m_1$ and $m_2$ by the first normalization scheme. The LLSS fit values obtained for $m_1$ and $m_2$ were 1992 and 1.0012 respectively, to be compared with the exact values 2000 and 1.000.

**Figure 1:**

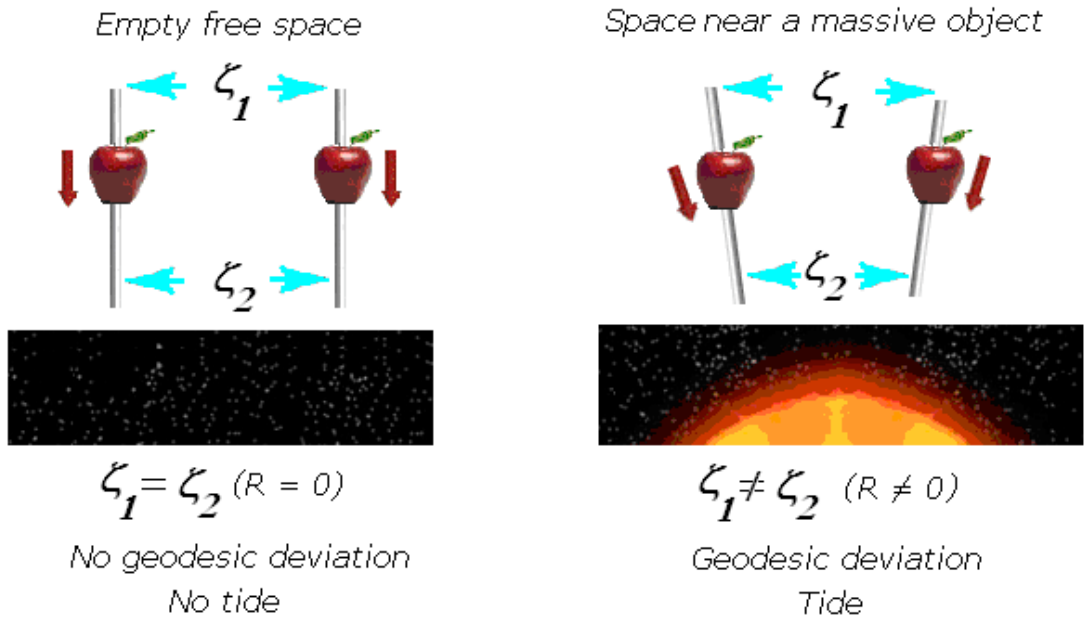

**Figure 2:**

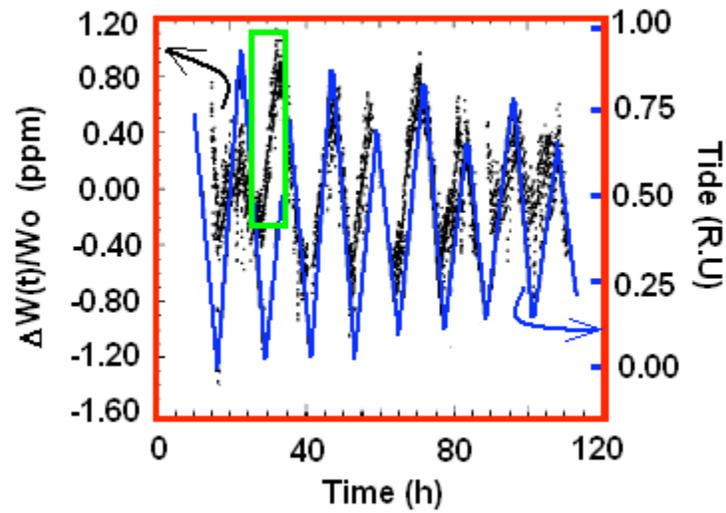

**Figure 3:**

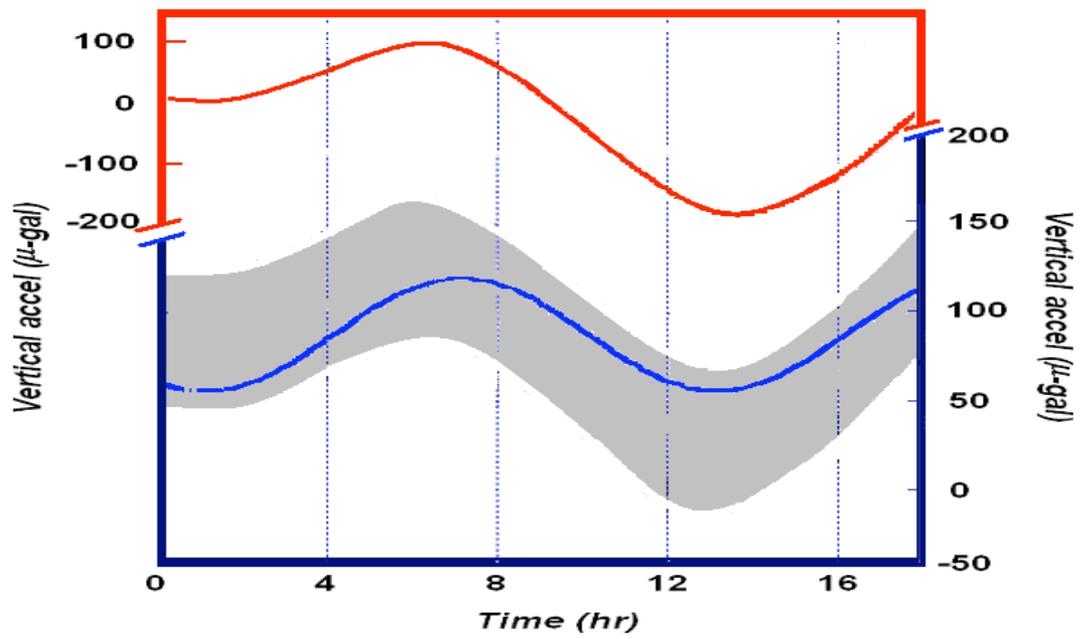

**Figure 4:**

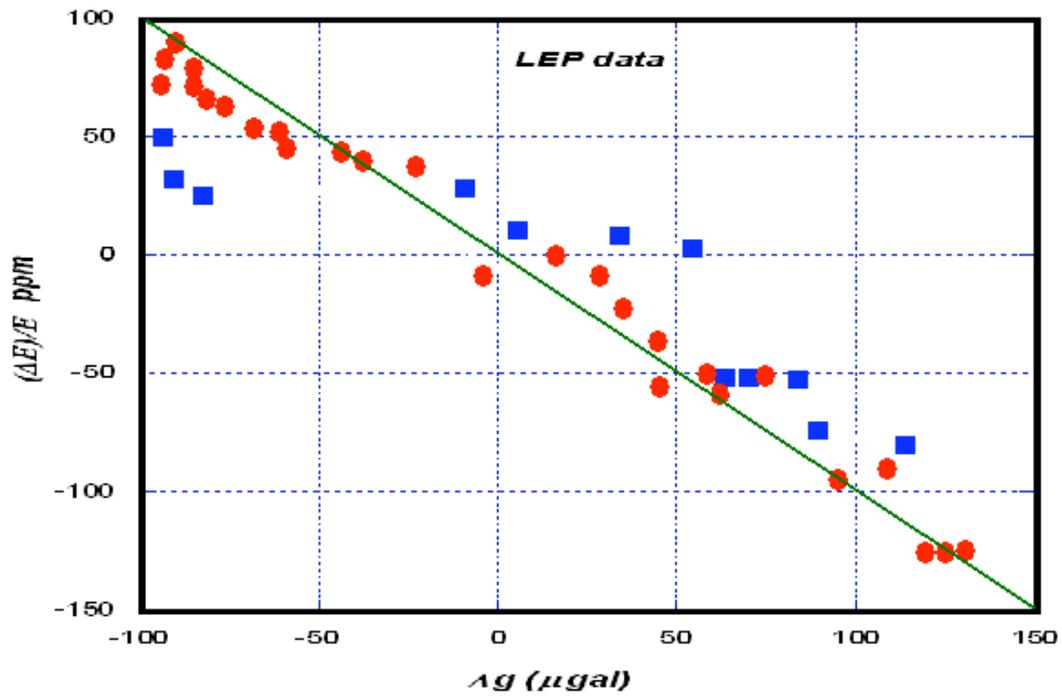

**Figure 5:**

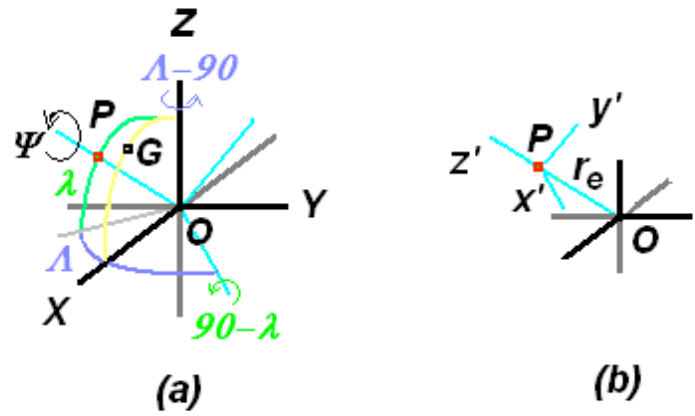

**Figure 6:**

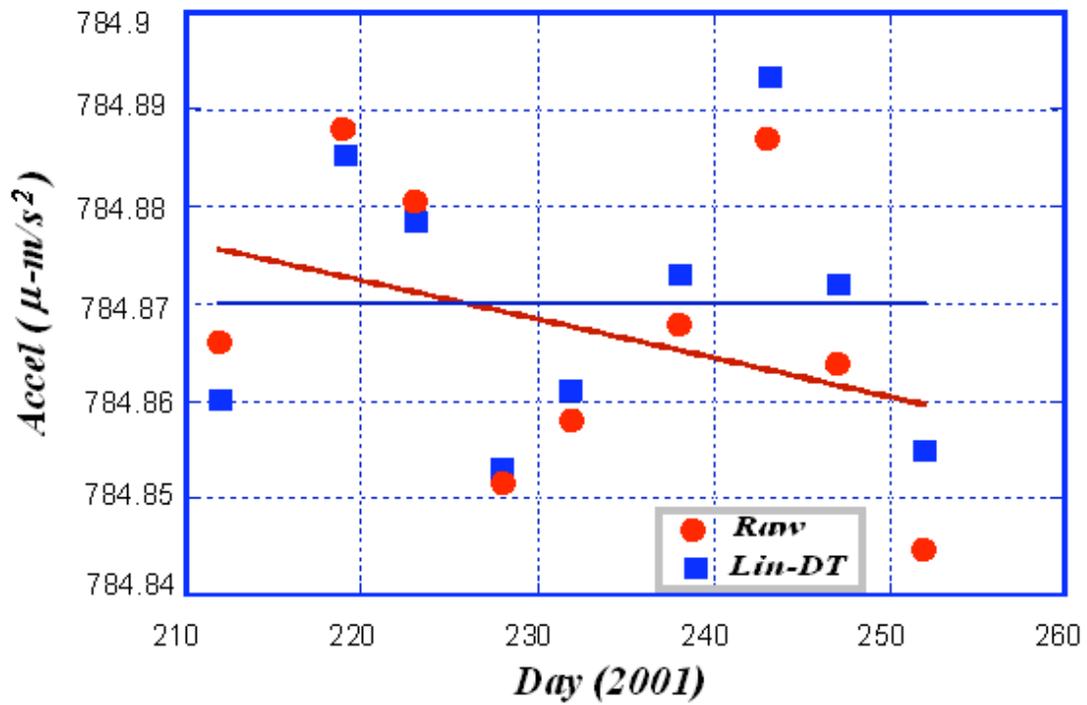

**Figure 7:**

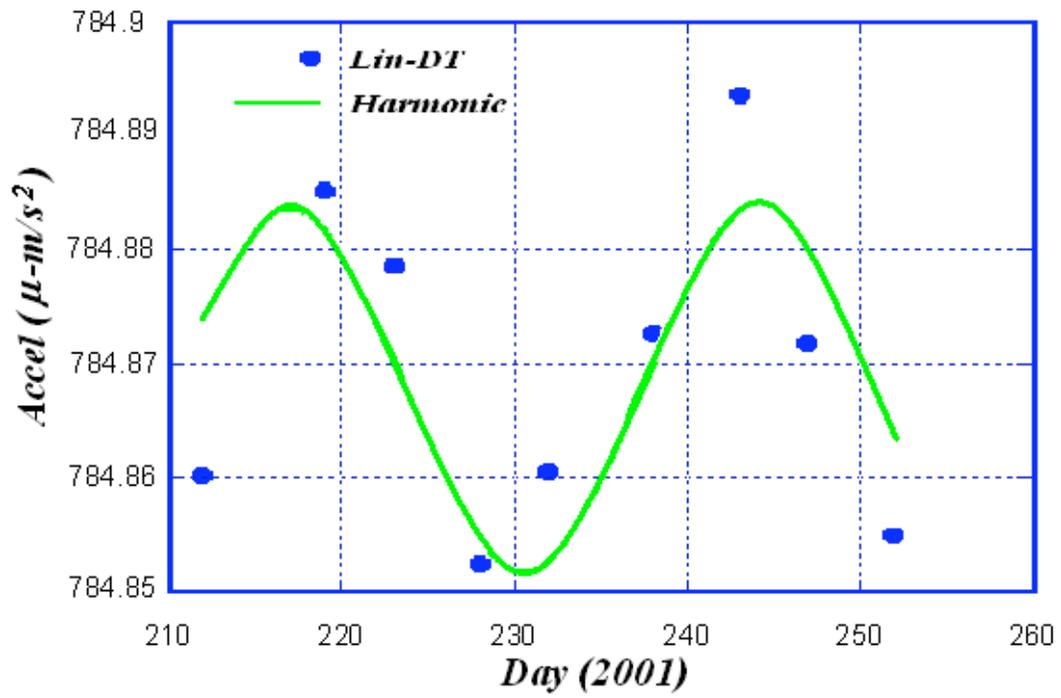

**Figure 8:**

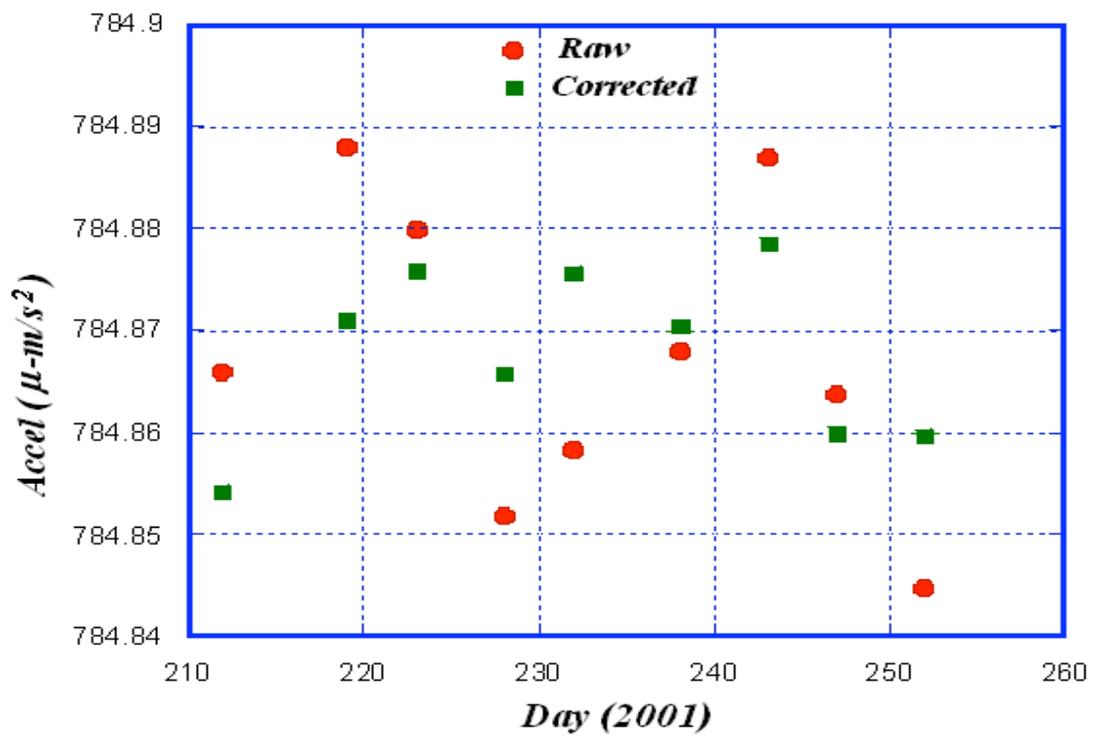

**Figure 9:**

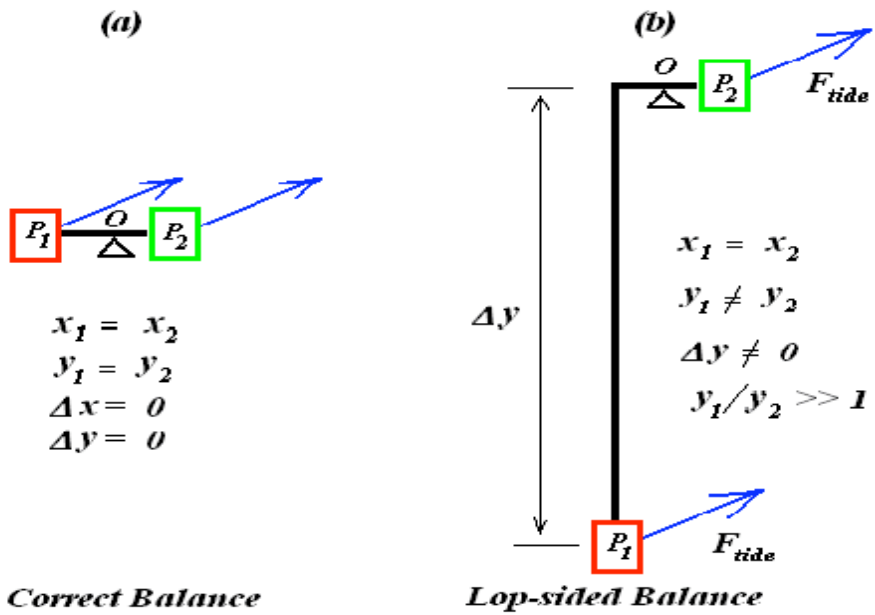

*(a)*

$x_1 = x_2$
$y_1 = y_2$
$\Delta x = 0$
$\Delta y = 0$

**Correct Balance**

*(b)*

$F_{tide}$

$\Delta y$

$x_1 = x_2$
$y_1 \neq y_2$
$\Delta y \neq 0$
$y_1/y_2 \gg 1$

$F_{tide}$

**Lop-sided Balance**

**Figure 10:**

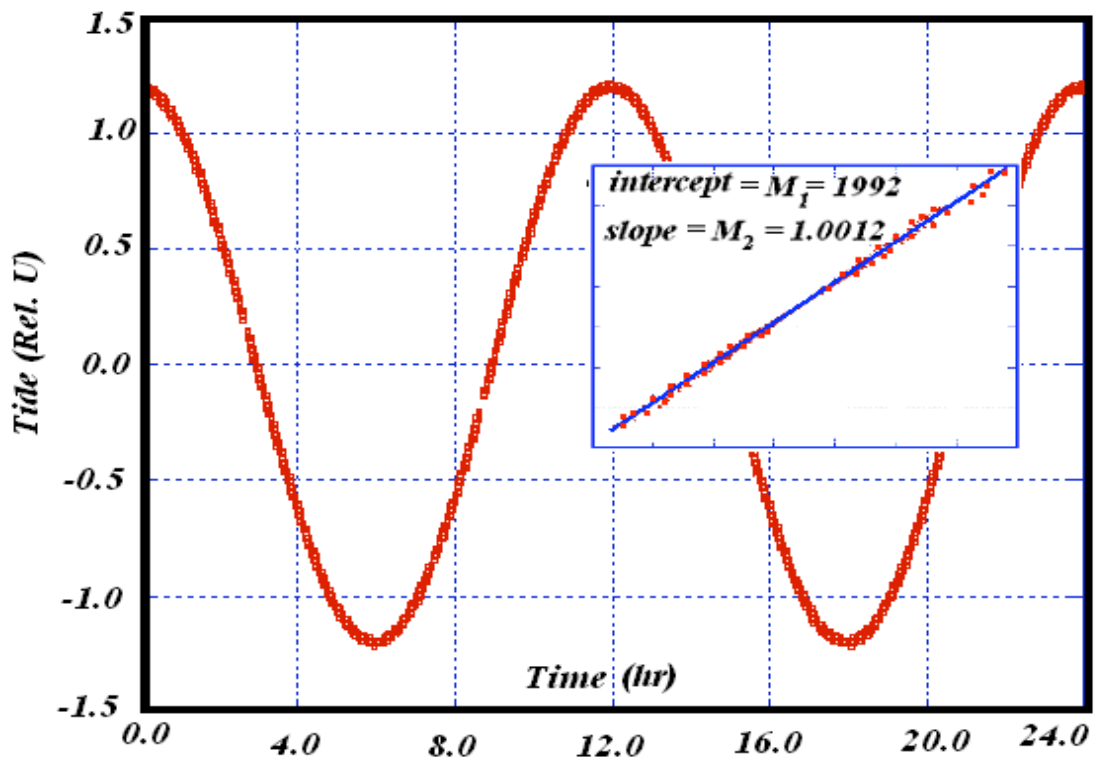

*intercept = $M_1$ = 1992*
*slope = $M_2$ = 1.0012*